\documentclass[pre,twocolumn,amsmath,amssymb,superscriptaddress,longbibliography,showpacs,nofootinbib]{revtex4-1}

\usepackage{color}
\usepackage{latexsym}

\usepackage{bm, mathtools}
\usepackage{graphicx}% Include figure files
\usepackage{graphics,epsfig,color}
\usepackage{dcolumn}% Align table columns on decimal point
\usepackage{times}
\usepackage[english]{babel}
\usepackage[latin1]{inputenc}
\usepackage[T1]{fontenc}

\newcommand{\dd}{{\rm d}}
\newcommand{\mean}[1]{\left< #1 \right>}
\newcommand{\Obs}{\mathcal{O}}
\renewcommand{\ge}{\geqslant}
\renewcommand{\le}{\leqslant}
\newcommand{\ie}{{\it i.e}}

\newcommand{\ee}{e}

\newcommand{\del}{\partial}

\newcommand{\kB}{k_{\rm B}}
\newcommand{\Eqref}[1]{Eq.~\eqref{#1}}
\newcommand{\genL}{{\mathbb L}}
\renewcommand{\vec}{\bm}
\newcommand{\Cp}{\mathcal{C}}
\definecolor{linkcolor}{rgb}{0,0,0.6} %hyperlink
\usepackage[pdftex,colorlinks=true, pdfstartview=FitV, linkcolor= linkcolor, citecolor= linkcolor, urlcolor= linkcolor, hyperindex=true,hyperfigures=true]{hyperref} %hyperlink% or \documentclass[page-classic]{epl2} for one column style

\begin{document}

\title{Thermal response of nonequilibrium RC-circuits}

\author{Marco Baiesi}
\email{baiesi@pd.infn.it}
\affiliation{
Department of Physics and Astronomy, University of Padova, 
Via Marzolo 8, I-35131 Padova, Italy
}
\affiliation{
INFN, Sezione di Padova, Via Marzolo 8, I-35131 Padova, Italy
}

\author{Sergio Ciliberto}
\affiliation{Laboratoire de Physique de Ecole Normale Supérieure de Lyon, 46 Allée d'Italie, 69364  Lyon, 
France}

\author{Gianmaria Falasco}
\affiliation{Max Planck Institute for Mathematics in the Sciences, Inselstr. 22,
04103 Leipzig, Germany}

\author{Cem Yolcu}
\email{yolcu@pd.infn.it}
\affiliation{
Department of Physics and Astronomy, University of Padova, 
Via Marzolo 8, I-35131 Padova, Italy
}

\date{\today}

\begin{abstract}
We analyze experimental data obtained from an electrical circuit
having components at different temperatures, showing how to predict
its response to temperature variations. This illustrates in detail how
to utilize a recent linear response theory for nonequilibrium
overdamped stochastic systems.  To validate these results, we
introduce a reweighting procedure that mimics the actual realization
of the perturbation and allows extracting the susceptibility of the
system from steady state data. This procedure is closely related to
other fluctuation-response relations based on the knowledge of the
steady state probability distribution.  As an example, we show
that the nonequilibrium heat capacity in general does not correspond to the
correlation between the energy of the system and the heat flowing into
it.  Rather, also non-dissipative aspects are relevant in the
nonequilbrium fluctuation response relations.
\end{abstract}


\maketitle

\section*{Introduction}

Understanding how a system responds to variations of its parameters is
one of the basic features of science. It is well
known that systems in thermodynamic equilibrium when slightly perturbed 
find their way back to a new steady regime by dissipation. The spontaneous
correlations in the unperturbed system
between this transient entropic change and an
observable anticipates us how that observable would
react to the actual perturbation. This is at the basis of the
fluctuation-dissipation theorem and of related response relations,
which hold in great generality in equilibrium~\cite{tod92,mar08}.

Out of equilibrium, in contrast, there are multiple linear response theories~\cite{mar08,bai13,bas15}, 
some based on the manipulation of the density of states \cite{aga72,fal90,spe09,sei10,pro09}, 
some on dynamical systems techniques for evolving observables~\cite{rue09,col12,col14}, 
and some on a path-weight approach for stochastic systems~\cite{cug94,che08,ver11,lip05,bai09}. 
The latter has revealed that entropy production is not sufficient for understanding 
the linear response of nonequilibrium systems. There are rather non-dissipative aspects of the
system vs perturbation relation that are equally relevant.

Within the linear response theory one finds recent approaches focusing on temperature 
perturbations~\cite{che09b,bok11,bai14,bra15, for15, pro15,yol16,fal16,fal16b}, which
lead for instance to a formulation of nonequilibrium heat capacity~\cite{bok11}, a notion
that should be useful for constructing  a steady state thermodynamics~\cite{sek07,sag11,ber13,mae14,kom15}.
The question is how a system far from equilibrium reacts to a change of one
or many of its bath temperatures. For example, one could be interested in
the response to temperature variations of a glassy system undergoing a relaxation process~\cite{mau10,gom12}.
Alternatively, a nonequilibrium steady state may be imposed by putting the system in contact
with two reservoirs at different temperatures~\cite{cil13,ber14}. It is the case of an experiment recently
realized with a simple desktop electric circuit in which one resistor was kept at room
temperature while the other was maintained at a lower temperature~\cite{cil13}.

In this paper, we analyze the experiments of the thermally unbalanced electric circuit~\cite{cil13}
and we discuss the good performances of a fluctuation-response relation~\cite{fal16,fal16b} in computing
the susceptibility of the system to a change of the colder, manipulable temperature.
The primary goal of this work is to show how to apply this approach in practice.
This is a stand-alone procedure for predicting the thermal linear response of the system.
Just to validate its results, we compare them with an alternative estimate of the
susceptibility, which is introduced here to exploit the knowledge of the steady state data
(which is accessible for the simple system analyzed), 
used by us in a reweighted form to replace the actual application of the perturbation. 
This useful procedure, as a byproduct, constitutes a new result of this work. 
We also show the connection of this reweighting procedure with another fluctuation-response
relation based on the steady state distribution, put forward by Seifert and Speck~\cite{sei10}.

In the following section we describe the experimental setup, then we recall the
structure of the fluctuation-response relation and we specialize it to our system.
In Sec.~\ref{sec:rewe} we introduce the reweighting procedure, and in Sec.~\ref{sec:res}
we show how to compute a nonequilibrium version of the heat capacity.
The conclusions are followed by an appendix in which we recall in detail the steps to
compute the Gaussian steady state distribution of linear stable systems and we
specify its form for the electrical circuit.

%%%%%%%%%%%%%%%%%%%
\begin{figure}[t!]
  \centering
  %\includegraphics[width=0.4\textwidth]{circuit_heat_exchange.pdf}
  %includegraphics[width=0.4\textwidth]{spring_particles.pdf}
  \includegraphics[width=0.4\textwidth]{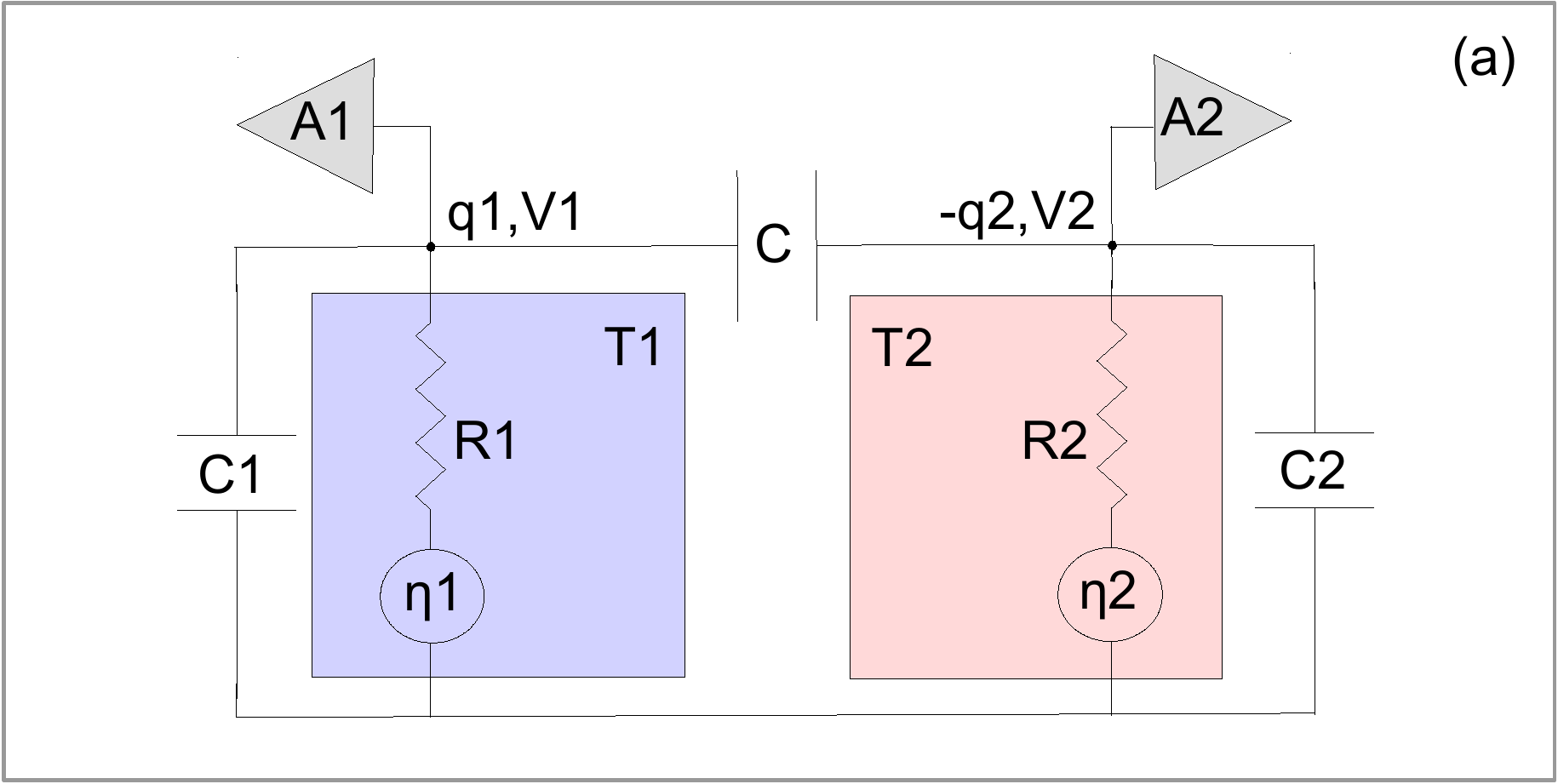}
  \includegraphics[width=0.4\textwidth]{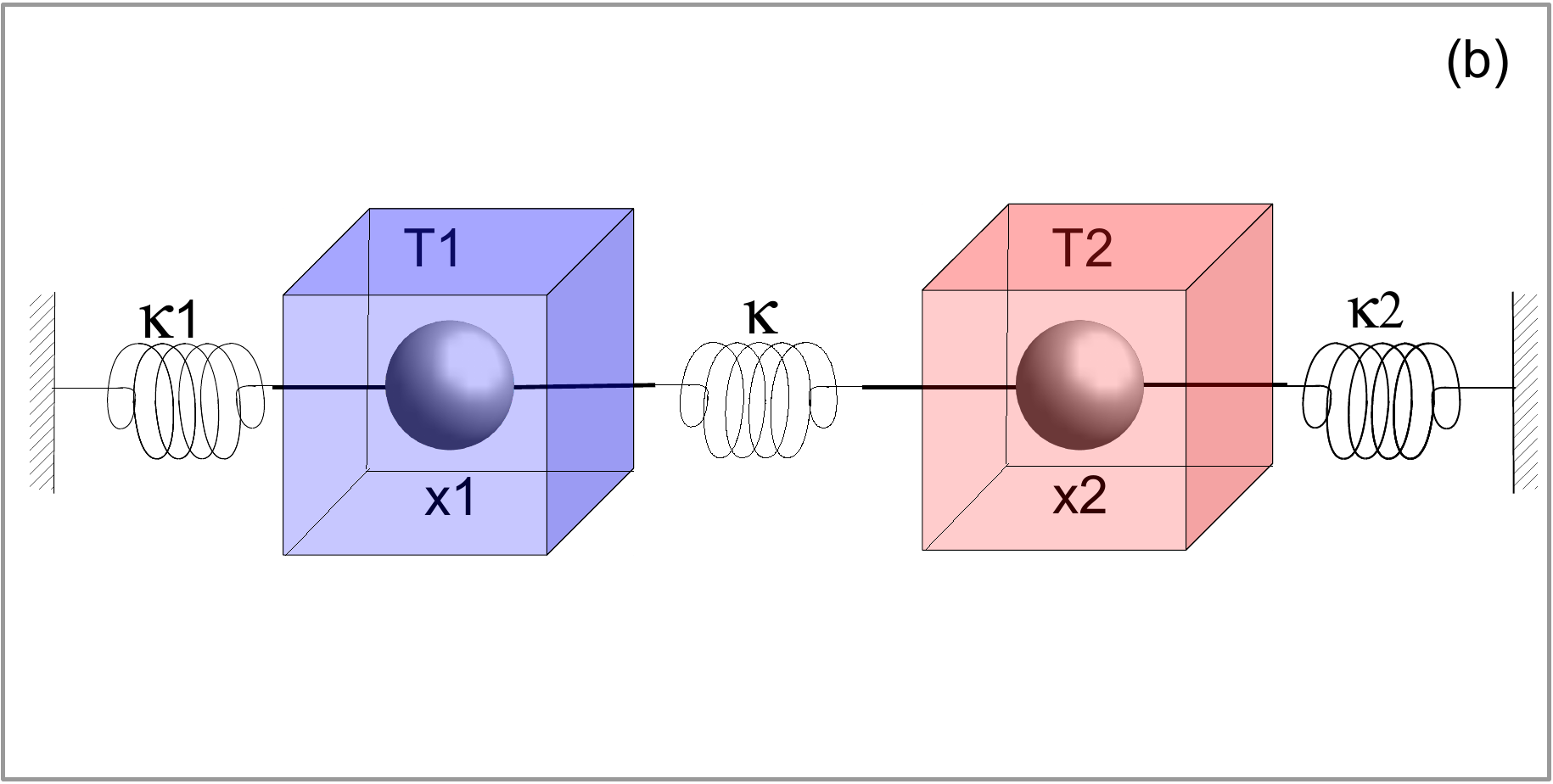}
  \caption{(Color online) (a) Diagram of the circuit. The resistor $R_1$ is kept at
    temperature $T_1$ while $R_2$ is always at room temperature
    $T_2=296K$. They are coupled via the capacitor $C$. The capacitors
    $C_1$ and $C_2$ account for the capacitance of the wiring,
    etc. The voltage generators $\eta_1$ and $\eta_2$ represent
    thermal fluctuations of the voltage that the resistors
    undergo. (b) Equivalent mechanical system with two Brownian
    particles moving in fluids at different temperatures $T_1$ and
    $T_2$, but trapped and coupled by harmonic springs. }
\label{fig:circuit}
\end{figure}
%%%%%%%%%%%%%%%%%%%

\section{Experimental set-up }\label{section_exp}
Our experimental set-up is sketched in Fig.\ref{fig:circuit}(a). It is constituted by two resistors $R_1$ and $R_2$, which are kept at different temperature, $T_1$ and $T_2$, respectively. These temperatures are controlled  by thermal baths and $T_2$ is kept fixed at $296K$ whereas $T_1$  can be set at a value  between $296K$ and $88K$  
using the stratified vapor above a liquid nitrogen bath.  
   The coupling capacitor $C$ controls the electrical power exchanged between the resistors  and as a consequence the energy exchanged between the two baths. No other coupling exists between the two resistors, which are inside two separated screened boxes. 
The quantities $C_1$ and $C_2$ are the capacitances of the circuits and the cables. 
Two extremely  low noise amplifiers  $A_1$ and $A_2$ \cite{can09} measure the voltage $V_1$ and $V_2$ across  the resistors $R_1$ and $R_2$, respectively. All the relevant quantities considered in this paper can be derived by the measurements of $V_1$ and $V_2$, as discussed below. 
In particular, the relationships between the measured voltages and the charges are 
\begin{eqnarray}
q_1&=& (V_1-V_2) \, C + V_1\, C_1  \label{eq_q1} \, , \\
q_2&=& (V_1-V_2) \, C - V_2\, C_2  \label{eq_q2} \, .
\end{eqnarray}
Assuming an initially neutral circuit, we denote by $q_1$ the
charge that has flown through the resistor $R_1$ into the node at
potential $V_1$, and by $q_2$ the charge that has flown through $R_2$
\emph{out of} the node at $V_2$. By analyzing the circuit one finds
that the equations of motion for these charges are
\begin{subequations}
\label{q-dyn}
 \begin{align}
 R_1 \dot q_1 &= - {C_2 \over X} q_1+ {C \over X} (q_2-q_1) + \eta_1 \\
 R_2 \dot q_2 &= - {C_1 \over X} q_2+ {C \over X} (q_1-q_2) + \eta_2 
\end{align}
\end{subequations}
where 
\begin{align}
\label{X}
X=C\, C_1  +C\, C_2+C_1\, C_2
\end{align}
and $\eta_i (t)$ is a white noise satisfying $\mean{\eta_i(t)\eta_j(t')}=2 \delta_{ij} {\kB T_i R_i} \delta(t-t')$.
Indeed, in Fig.\ref{fig:circuit}(a) the two resistances have been drawn with their associated thermal noise generators $\eta_1$ and $\eta_2$, whose power spectral densities  are given by the Nyquist formula $|\tilde \eta_m|^2= 4 k_B R_mT_m$, with $m=1,2$.

More details on the experimental set-up can be found in Ref.~\cite{cil13}. For the data used for the analysis discussed in the following section, the values of the components are:   $C=100$pF,  $C_1=680$pF, $C_2=420$pF and $R_1=R_2=10$M$\Omega$. The longest characteristic time of the system is 
$Y={(C_1+C) R_1 +(C_2+C) R_2 }$, which for the mentioned values of the parameters is $Y=13\,$ms.

\section{Thermal response}

The system has $N=2$ degrees of freedom.
Eqs.~\eqref{q-dyn} can be mapped onto the mechanical system in
Fig.\ \ref{fig:circuit}(b) involving two Brownian particles coupled by
harmonic springs,
\begin{subequations}
\label{x-dyn}
 \begin{eqnarray}
\dot x_1&=& \mu_1 F_1(\vec x) + \sqrt{2 \mu_1 \kB  T_1}\, \xi_1  \\
\dot x_2&=& \mu_2 F_2(\vec x) + \sqrt{2 \mu_2 \kB  T_2}\, \xi_2 
\end{eqnarray}
\end{subequations}
Here $\vec x = (x_1,x_2)$ are the two positions with $x_i=0$ when the
springs are at rest, $\vec T = (T_1,T_2)$ the temperatures, and the
(harmonic) forces $\vec F=(F_1,F_2)$ are derived from the
potential
\begin{align}
  U(\vec x) = \frac 1 2 \left[\kappa_1 x_1^2 + \kappa (x_2-x_1)^2 +
    \kappa_2 x_2^2\right] \ . \label{eq:U}
\end{align}
The detailed mapping between the electrical and mechanical models is
summarized in Table~\ref{tab:1}; for instance, the admittance $1/R_1$
is mapped to the mobility $\mu_1$.  Again, each Gaussian white noise
$\xi_j$ is uncorrelated from the other,
\begin{align}\label{noise}
\mean{\xi_j(t)\xi_{j'}(t')} =  \delta(t-t') \delta_{jj'} .
\end{align}

This recasting in the form \eqref{x-dyn} allows us to use some
recently introduced thermal response formulas~\cite{fal16,fal16b}.
They predict the linear response of an overdamped stochastic system
with additive noise, in general nonequilibrium conditions, when the
perturbation is a change of one or more temperatures. In accordance
with the presentation of that approach, we choose natural units
($\kB=1$) in the following, taking temperatures to have dimensions of
energy. %% drop the Boltzmann constant
%% from the formulas and we convert temperatures from kelvins to joules
%% times the value of $\kB $.

%%%%%%%%%%%%%%%%%%%%%%%%%%%%%%%%%%%%%%%%%%%%%%%%%%%%%%%%%%%%%%%%%%% \multicolumn{2}{c|}{model B}
\begin{table}[!b]
\begin{center}
\begin{ruledtabular}
\begin{tabular}{ l  l }
electrical & mechanical  \\
\hline
$q_1$ & $x_1$ \\
$q_2$ & $x_2$ \\
$1/ R_1$ & $\mu_1$ (mobility)\\
$1/ R_2$ & $\mu_2$ \\
$C_2/X$  & $\kappa_1$ (spring constant)\\  
$C_1/X$  & $\kappa_2$ \\  
$C/X$    & $\kappa$ \\
%% \hline
%% & $F_1(\vec x) = -\kappa_1 x_1 + \kappa (x_2 - x_1) $\\
%% & $F_2(\vec x) = -\kappa_2 x_2 + \kappa (x_1 - x_2) $\\
%% \hline
%% & $\partial_1 F_1(\vec x) = -\kappa_1 - \kappa $ \\
%% & $\partial_2 F_1(\vec x) = \kappa $ \\
%% \hline
%% & $\genL^{(T_1)} F_1(\vec x) = \mu_1 F_1(\vec x) (-\kappa_1 - \kappa) $ \\
%% & $\genL^{(T_2)} F_1(\vec x) =  \frac{T_1}{T_2} \mu_2 F_2(\vec x) \kappa $ \\
\end{tabular}
\end{ruledtabular}
\end{center}
\caption{Mapping between electric quantities and mechanical ones. Note
  the inversion of indices for $C_2/X \to \kappa_1$ and $C_1/X \to
  \kappa_2$.}
\label{tab:1}
\end{table}
%%%%%%%%%%%%%%%%%%%%%%%%%%%%%%%%%%%%%%%%%%%%%%%%%%%%%%%%%%%%%%%%%%%

The thermal susceptibility of a state observable $\Obs(\vec x)$ is
defined as the response to a step variation $\vec T \to \vec \Theta$
of the set of temperatures, parametrized by a function $\theta(t)=0$
for times $t< 0$ and $\theta(t)=$constant for $t\ge 0$.  In
particular, with indicators $\epsilon_i$ ($1\le i\le N$) that specify
which temperatures receive the perturbation, here we write
\begin{align}
\vec \Theta \equiv (T_1+\epsilon_1 \theta,T_2+\epsilon_2 \theta) \ .
\end{align}
The susceptibility as a function of time $t$ is then
\begin{align}
  \chi^\theta_\Obs (t) = \lim_{\theta \to 0} 
  \frac {\left< \Obs(\vec x(t)) \right>_{\vec T,\vec \Theta} 
    - \left< \Obs(\vec x(t)) \right>_{\vec T,\vec T}} {\theta} \ .
\label{chi1}
\end{align}
In the averages, the first subscript represents the initial ($t<0$)
temperatures, while the second subscript represents 
the temperatures under which the observed dynamics ($t\ge 0$) takes place.

A recent fluctuation-response relation~\cite{fal16,fal16b} 
expresses the susceptibility \eqref{chi1} of the state observable $\Obs(\vec x)$
as a sum,
\begin{align}
\label{chi}
\chi_{\Obs}(t) = S_1 + S_2 + K_1 + K_2
\end{align}
where the terms are
\begin{subequations}
\label{SSKK}
%\begin{widetext}
\begin{align}
\label{chi_S1}
S_1\!=& -  \mean{  \Obs (t)\, 
\sum_i \frac{\epsilon_i}{2 T_i^2}\int_0^t \dot x_i(t') F_i(t')  \, \dd t'} 
\\
\label{chi_S2}
S_2\!=& 
\mean{ \Obs (t) 
\sum_i \frac{\epsilon_i}{4 T_i^2} 
\sum_{j=1}^N \left(\frac{T_i}{T_j}-1 \right) \int_0^t  [x_i \dot x_j \partial_j F_i](t') \, \dd t'} 
\\
\label{chi_K1}
K_1\!=&  
\Bigg< \Obs(t) \sum_i \frac{\epsilon_i}{4 T_i^2} 
\int_0^t \left[ \mu_i F_i^2 + x_i  \genL^{(T_i)} F_i \right]\!(t') \, \dd t' \Bigg>
\\
\label{chi_K2}
K_2\!=& 
\left.\frac{\dd}{\dd t'}\mean{\Obs(t) 
\sum_i \frac{\epsilon_i}{8 \mu_i T_i^2} x_i^2(t')}\right|_{t'=0}^{t'=t}
\end{align}
%\end{widetext}
\end{subequations}
with the shorthand $\partial_j = \partial/\partial x_j$, and
$\mean{\ldots} = \mean{\ldots}_{\vec T,\vec T}$ denoting unperturbed averages
which have an understood dependence on the distribution $\rho_0(\vec
x(0))$ at the time when the perturbation is turned on.  (Let us stress
that the labels $1,2$ of these $S$ and $K$ terms have nothing to do
with the index of the resistors, particles, etc.)  Integrals are in the
Stratonovich sense, hence in their discretized version one performs
midpoint averages, such as 
$\dot x(t) F(t) \dd t \to [x(t+\dd t) -  x(t)]\frac 1 2 [F(t+\dd t)+F(t)]$.  
(However, temperatures and mobilities do not depend on the coordinates 
and the interpretation of the stochastic equation is free.) 

The term $S_1$ is a standard correlation between observable and
entropy production, but it contains a prefactor $1/2$ not present
in the equilibrium version (Kubo formula).  The term $S_2$ instead
correlates the observable with another form of entropy production and
clearly it is relevant only if $T_j \ne T_i$ for some $(i,j)$.  The
terms $K_1$ and $K_2$, previously called the {\em frenetic} terms~\cite{bai13,bai14,yol16,fal16,fal16b},
instead correlate the observable with time-symmetric aspects of the
dynamics. These are necessarily non-dissipative in nature.  In all
cases it is understood that we are dealing with quantities in excess
due to the perturbation. The generalized generator
\begin{equation}
\genL^{(T_i)} = \sum_j \frac {T_i}{T_j} \left[ \mu_j F_j(\vec x) \partial_{j} + \mu_jT_j \partial^2_{j}\right]
\end{equation}
was introduced to describe the evolution of the degrees of freedom in
terms of the $j$-th {\it thermal time} as dictated by the $i$-th
temperature (see~\cite{fal16b} for more details). 
It differs from the backward generator of the dynamics~\eqref{x-dyn},
\begin{equation}
\genL = \sum_{j=1}^N \left[ \mu_j F_j(\vec x) \partial_{j} + \mu_jT_j \partial^2_{j}\right]\, ,
\end{equation}
whose action on a state function inside an average is expressed as 
$\frac{\dd}{\dd t} \mean{\Obs(\vec x(t))}= \mean{\genL \Obs(\vec x(t))}$.  
The definition of a thermal time permits to recast \eqref{x-dyn} as isothermal dynamics. 
For example, if $i=1$ and hence $T_1$ is taken as a reference,
then the thermal time $\tau_2= t T_2/T_1 $ yields for $x_2$
\begin{align} 
\frac{d x_2}{d \tau_2}&= \frac{T_1}{T_2} \mu_2 F_2(\vec x) + \sqrt{2 \mu_2 \kB  T_1}\, \xi_2(\tau_2),
\end{align}
where the different intensity of the noise $\xi_2$ (it has now $T_1$ in the prefactor) 
is associated with a rescaling $\sim T_1/T_2$ of the mechanical force $F_2$.

In our analysis we work with experimental trajectory data collected in
steady states, where $d/dt' \mean{\Obs(t)x^2(t')} = -d/dt \mean{\Obs(t)x^2(t')} = -\mean{\genL \Obs(t)x^2(t')}$
for $t\ge t'$.
Hence we rather use the alternative form
\begin{align}
\label{chi_K2ss}
K_2^{s} & =   \mean{ \genL \Obs(t) \sum_i \frac{\epsilon_i}{8 \mu_i T_i^2}[x_i^2(0)-x_i^2(t)]} 
\end{align}
because it is numerically more stable than $K_2$~\cite{fal16b}. 
Since in the given experimental setup it is natural to manipulate
$T_1$ (while the room temperature, $T_2$, remains unperturbed), we
show examples with $\epsilon_1=1$ and $\epsilon_2=0$.  This leads to
the susceptibility $\chi_\Obs(t)$ being composed of the specific terms
\begin{subequations}
\label{chi_fr_2}
\begin{align}
\label{chi_S1b}
S_1 &= -  \frac{1}{2 T_1^2}  \mean{  \Obs (t)\, \int_0^t  \dot x_1(t') F_1(t') \, \dd t'} 
\\
\label{chi_S2b}
S_2 &= 
\frac{1}{4 T_1^2} 
\mean{ \Obs (t) 
\left(\frac{T_1}{T_2}-1 \right) \int_0^t  [x_1 \dot x_2 \partial_2 F_1](t') \, \dd t'}  
\\
\label{chi_K1b}
K_1 &=  \frac{1}{4 T_1^2} 
\mean{\Obs(t)
 \int_0^t \left[\mu_1 F_1^2  + x_1 \genL^{(T_1)} F_1 \right]\!(t') \, \dd t'} 
\\
\label{chi_K2b}
K_2 & =  \frac{1}{8 \mu_1 T_1^2} \mean{ \genL \Obs(t)  [x_1^2(0)-x_1^2(t)]} 
\end{align}
\end{subequations}
with $\genL^{(T_1)} = \mu_1 [ F_1 \partial_1 + T_1 \partial^2_1] +
\mu_2 [ \frac{T_1}{T_2} F_2 \partial_2 + T_1 \partial^2_2]$.  Note
that we have dropped the superscript ``s'' from $K_2$.

The susceptibility is found as the sum of these correlations with
fluctuating trajectory functionals, {\em predicting} the
susceptibility without actually performing perturbations. Before
showing examples, in the next section we describe a second procedure
aimed at computing the response in a more {\em direct} way. The latter
will then be compared with the fluctuation-response results above.

\section{Reweighting}
\label{sec:rewe}

In the analysis via the fluctuation-response relation exposed in the
previous section, we deal with experimental data collected in steady
states at various temperatures $T=(T_1,T_2)$. Next we show that the
same data can be used to extract a form of the susceptibility that is
equivalent to \Eqref{chi1}. This means that we can bypass once again the step of
the actual perturbation of the system in the laboratory.

In the definition \eqref{chi1} what is not useful is that one average is over
trajectories under the perturbed $\vec \Theta$, while the other is
over unperturbed trajectories. In steady state experiments,
trajectories of the former kind are not available.  To sidestep this,
we find it convenient to consider the alternative formula
\begin{align}
  \chi^\theta_\Obs (t) = \lim_{\theta\to 0} 
  \frac {\left< \Obs(\vec x(t)) \right>_{\vec \Theta,\vec T} 
    - \left< \Obs(\vec x(t)) \right>_{\vec \Theta,\vec \Theta}} {- \theta} \ ,
\label{eq:chi2}
\end{align}
because with this form, we can re-express both averages above
in terms of steady state averages at $\vec T$, by the following
arguments.

First, take the steady state average
\begin{align}
  \left< \Obs(\vec x(t)) \right>_{\vec \Theta, \vec \Theta} =&
  \left< \Obs(\vec x) \right>_{\vec \Theta, \vec \Theta} 
  = \int \dd \vec x \, \rho_{\vec \Theta} (\vec x) \Obs(\vec x) \\
  =& \int \dd \vec x \, \rho_{\vec T} (\vec x) 
  \frac {\rho_{\vec \Theta} (\vec x)} { \rho_{\vec T} (\vec x) }
  \Obs(\vec x) \nonumber\\
  =& \left< \frac {\rho_{\vec \Theta} (\vec x)} 
  { \rho_{\vec T} (\vec x)} \Obs(\vec x) \right>_{\vec T,\vec T} \ .
\end{align}
Second, by denoting the probability measure of path $[\vec x]$ under
temperatures $\vec T$ by $\mathcal{D} \vec x P_{\vec T} [\vec x]$
[where $ P_{\vec T} [\vec x]$ is the path-weight, given that it
starts from $\vec x(0)$],
take the transient average
\begin{align}
  \left< \Obs(\vec x(t)) \right>_{\vec \Theta,\vec T} =& 
  \int \mathcal{D} \vec x \, P_{\vec T} [\vec x] \rho_{\vec \Theta} (\vec x(0))
  \Obs(\vec x(t)) \\
  =& \int \mathcal{D} \vec x \, P_{\vec T} [\vec x] \rho_{\vec T} (\vec x(0))
  \frac {\rho_{\vec \Theta} (\vec x(0))} {\rho_{\vec T} (\vec x(0))}
  \Obs(\vec x(t)) \\  
  =& \left<  
  \frac {\rho_{\vec \Theta} (\vec x(0))} {\rho_{\vec T} (\vec x(0))}
  \Obs (\vec x(t)) \right>_{{\vec T},{\vec T}} \ .
\end{align}
Thus, by this reweighting via stationary distributions, both averages
appearing in \Eqref{eq:chi2} have been reformulated as steady state
averages at $\vec T$, and the susceptibility becomes
\begin{align}
  \chi^\theta_\Obs  (t) = 
  \lim_{\theta\to 0}  \frac {-1}{\theta}
  \bigg( 
  & 
  \left< \frac {\rho_{\vec \Theta} (\vec x(0))} {\rho_{\vec T} (\vec x(0))} \Obs (\vec x(t)) \right>_{{\vec T},{\vec T}} 
  \nonumber\\
  & 
  - \left< \frac {\rho_{\vec \Theta} (\vec x)} 
  { \rho_{\vec T} (\vec x)} \Obs(\vec x) \right>_{{\vec T},{\vec T}} \bigg) \ .
\end{align}
The second single-time average can be written at any instant of time
due to time-translation invariance. As such, substituting the
particular points $\vec x(0)$ or $\vec x(t)$ one obtains,
respectively,
\begin{align}
  \chi^\theta_\Obs  (t) = \lim_{\theta\to 0}  \frac {-1}{\theta} \left<
  \left[ \Obs(\vec x(t)) - \Obs(\vec x(0)) \right]
  \frac {\rho_{\vec \Theta} (\vec x(0))} {\rho_{\vec T} (\vec x(0))}
  \right>
\label{chi_rewe1}
\end{align}
or
\begin{align}
  \chi^\theta_\Obs  (t) = \lim_{\theta\to 0}  \frac {1}{\theta} 
  \left< \Obs(\vec x(t)) \left[
    \frac {\rho_{\vec \Theta} (\vec x(t))} {\rho_{\vec T} (\vec x(t))}
   -\frac {\rho_{\vec \Theta} (\vec x(0))} {\rho_{\vec T} (\vec x(0))}
    \right] \right> \ .
\label{chi_rewe2}
\end{align}
Again, $\mean{\ldots}$ means the steady state average
$\mean{\ldots}_{{\vec T},{\vec T}}$ with the available data.  Both
formulas can be used to extract the response of the system to a step
change of temperature(s) performed at $t=0$. In our analysis we chose
to use \Eqref{chi_rewe1}.

It is interesting to connect these expressions with previous response
relations based on the knowledge of the steady state distribution. One
notes that in the limit $\theta \to 0$, the reweighting factor
\begin{align}
  \frac {\rho_{\vec \Theta}} {\rho_{\vec T}} \simeq
   \frac {\rho_{\vec T} + \theta \, \del_\theta \rho_{\vec \Theta}} {\rho_{\vec T}}
  = 1 + \theta \, \del_\theta  \ln \rho_{\vec \Theta} \ .
  \label{eq:pratio}
\end{align}
Substituting this limit, and dropping for simplicity the temperature
indices, the second expression
\eqref{chi_rewe2} for susceptibility above becomes
\begin{align}
  \chi_\Obs (t) = \left< \Obs(\vec x(t)) \left[
    \del_\theta \ln \rho (\vec x(t)) - \del_\theta \ln \rho (\vec x(0))
    \right] \right> \ ,
\end{align}
implying that it comes from a response function 
[$\chi_\Obs (t) = \int_0^t  \dd s \, R_\Obs(t-s)$]
\begin{align}
  R(t-s) = \frac{\dd}{\dd s} \left< \Obs(\vec x(t))
  \del_\theta \ln \rho (\vec x(s)) \right> \, .
\label{SpSe1}
\end{align}
Equivalently, defining the stochastic entropy $\mathcal{I} = -\ln \rho$,
\begin{align}
  R(t-s) = - \frac{\dd}{\dd s} \left< \Obs(\vec x(t))
  \del_\theta \mathcal{I} (\vec x(s)) \right> \ ,
\label{SpSe2}
\end{align}
which is Speck and Seifert's response formula~\cite{sei10} for steady
states, with the only difference that $\theta$ carries a physical
dimension while usually the perturbation was expressed in terms of a
dimensionless parameter $h$. 

While an analytical expression such as \eqref{SpSe2}
is more elegant than \eqref{chi_rewe1} or \eqref{chi_rewe2}, 
on the practical side the former may be less convenient.
First of all, an analytical expression for the stationary
distribution may not be known or calculable, in which case it must be
actually measured at two different temperatures and the $\theta$
derivative will have to be performed discretely, which is equivalent
to using the expressions prior to \Eqref{eq:pratio}. Secondly, even if
an analytical expression for the stationary distribution is available
(as it is for the present system of interest; details in the
appendix), its $\theta$ derivative might be too unwieldy to work with,
from an implementation point of view. 
A discrete approximation for the derivative, such as \eqref{chi_rewe1},
is simpler to handle. We have indeed followed this path, using analytical
expressions for the distributions $\rho_{\vec \Theta}$ and 
$\rho_{\vec  T}$, choosing $\vec \Theta = (T_1 + \theta, T_2)$ with 
$\theta =T_1/100$.

\section{Nonequilibrium heat capacity}\label{sec:res}

In this section we show the analysis of experimental
data, which show that the fluctuation-response relation 
$\chi_\Obs(t)= S_1 + S_2 + K_1 + K_2$ 
with terms listed in \eqref{chi_fr_2}, can
reliably compute the susceptibility of the system. 
The knowledge of the steady state distribution allows us to
use the formula introduced in \eqref{chi_rewe1}, and to compute the
response independently. The two versions turn out to yield results in
good agreement with each other.

The averages used to compute susceptibilities are performed over 
trajectories that extend over $\approx 50$ ms, with 
time steps of length $\Delta t=1/8192 \, {\rm s} \approx 0.122 \,{\rm ms}$.
Each trajectory is extracted by choosing a different starting point 
from the steady state sampling.  Three cases
are considered, one in the equilibrium condition $T_1 = T_2 = 296$~K 
($\approx 7\times 10^7$ trajectories) and two far from equilibrium,
$T_1 = 140$~K ($\approx 2.6\times 10^7$ trajectories) and $T_1 = 88$~K
($\approx 4.6\times 10^7$ trajectories).

As an observable $\Obs(\vec x)$ -- where we recall that $\vec x
= (x_1 , x_2) = (q_1,q_2)$ -- we consider the total electrostatic
energy of the system, \Eqref{eq:U}, in accord with the mapping of
Table \ref{tab:1} between electrical and mechanical quantities.  The
backward generator acting on this observable, $\genL\Obs$ appearing in
\eqref{chi_K2b}, becomes
\begin{align}
\genL U(\vec x) =& \kappa_1\mu_1 (F_1(\vec x) x_1 + T_1) + \kappa_2\mu_2 (F_2(\vec x) x_2 + T_2)\nonumber \\
                 & \kappa [\mu_1 (F_1(\vec x) (x_1-x_2) + T_1) \nonumber\\
                 & \quad + \mu_2(F_2(\vec x) (x_2-x_1) + T_2)]
\end{align}
where we remind that $\kB=1$ and temperatures have dimensions of
energy. The response of the energy to a change of temperature becomes
the nonequilibrium version of the heat capacity if $T_1\ne T_2$
(a different definition of heat capacity for nonequilibrium
systems can be found in Ref.~\cite{bok11}).
The following analysis confirms that in general this heat capacity
cannot be computed only from
the correlation between energy and heat flowing into the 
system~\cite{bok11,bai14,yol16,fal16,fal16b},
unless this is in equilibrium. 

\begin{figure}[!t]
$T_1 = 296$ K\\
\includegraphics[width=.9\columnwidth]{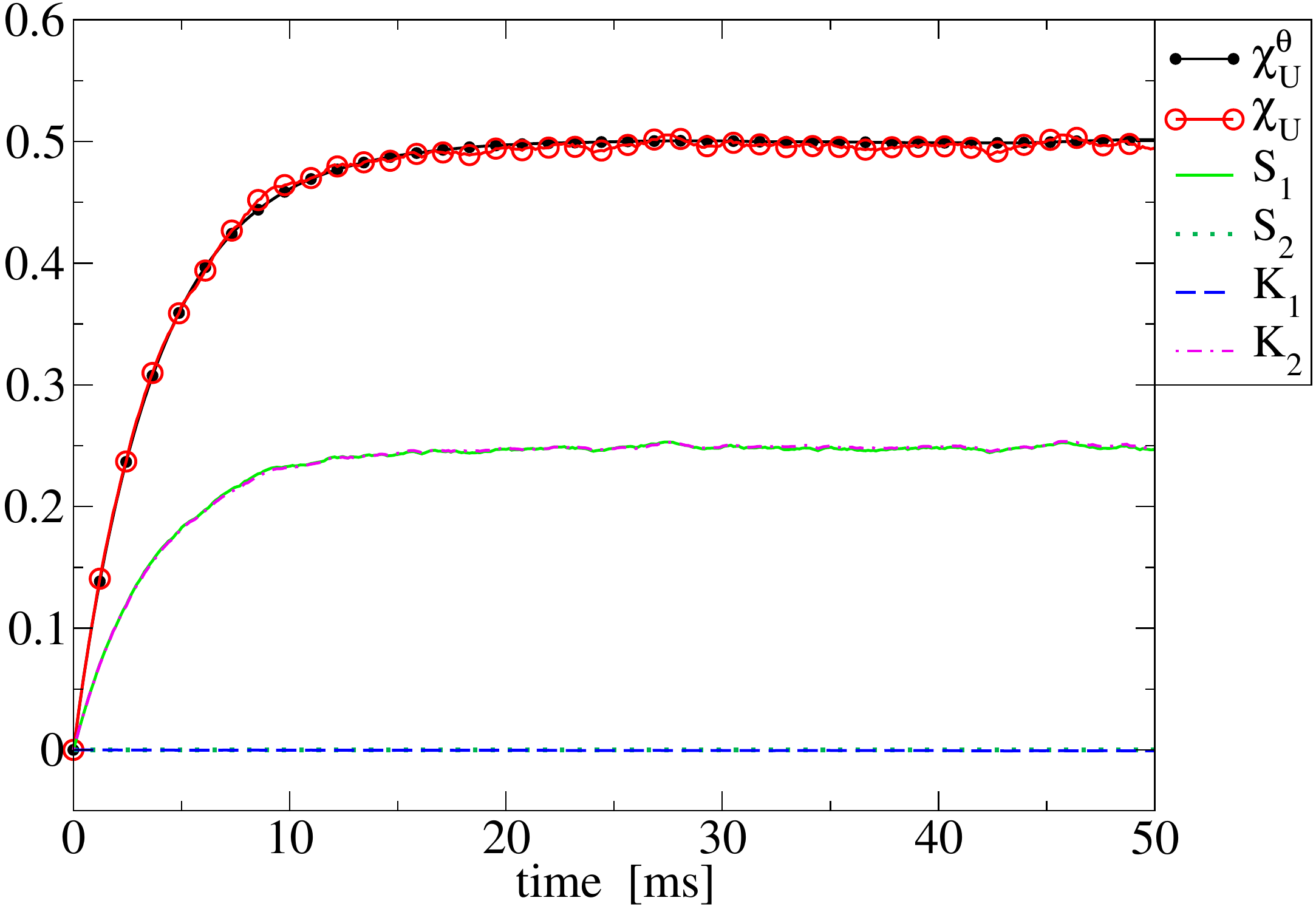}
$T_1 = 140$ K\\
\includegraphics[width=.9\columnwidth]{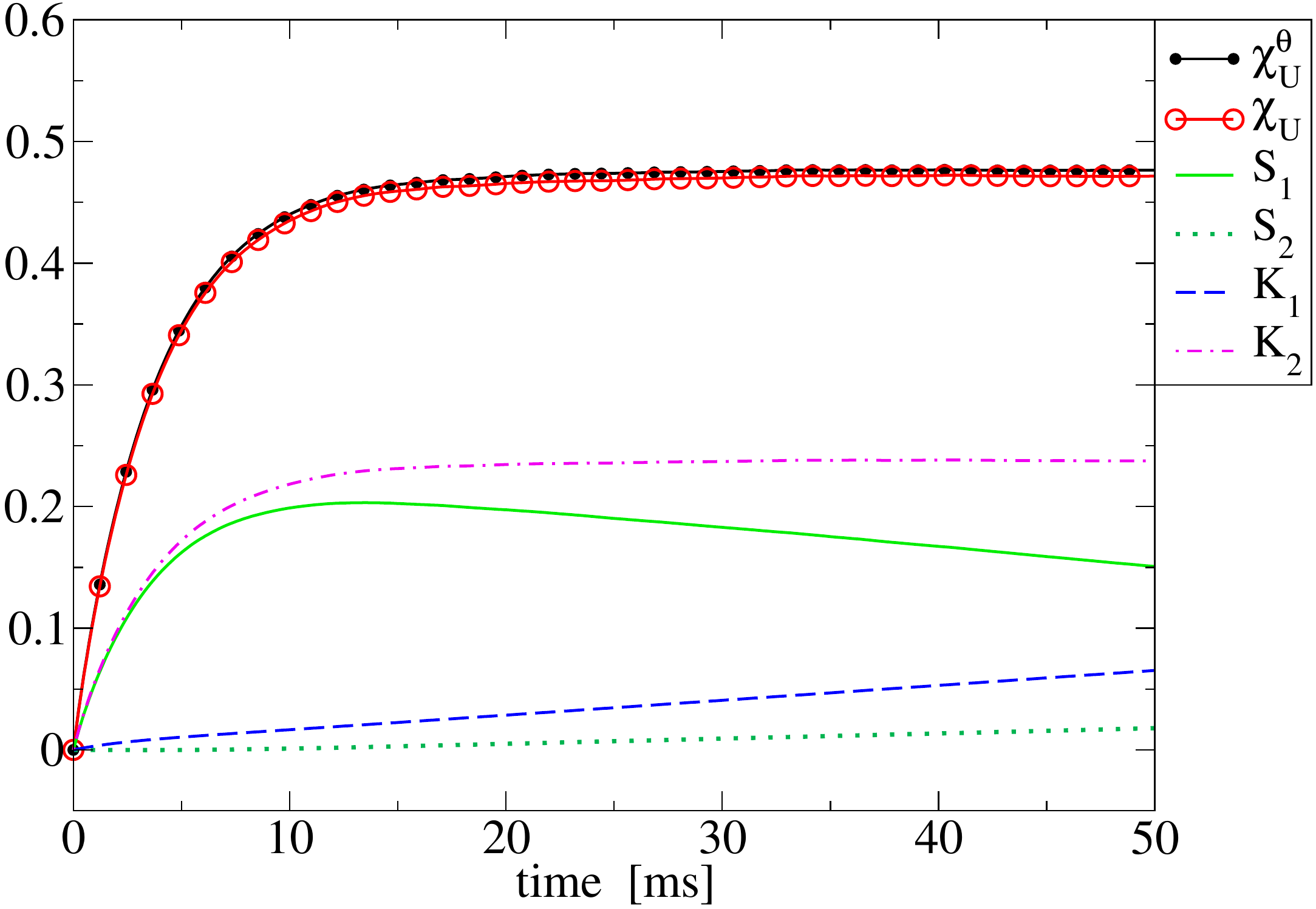}
$T_1 = 88$ K\\
\includegraphics[width=.9\columnwidth]{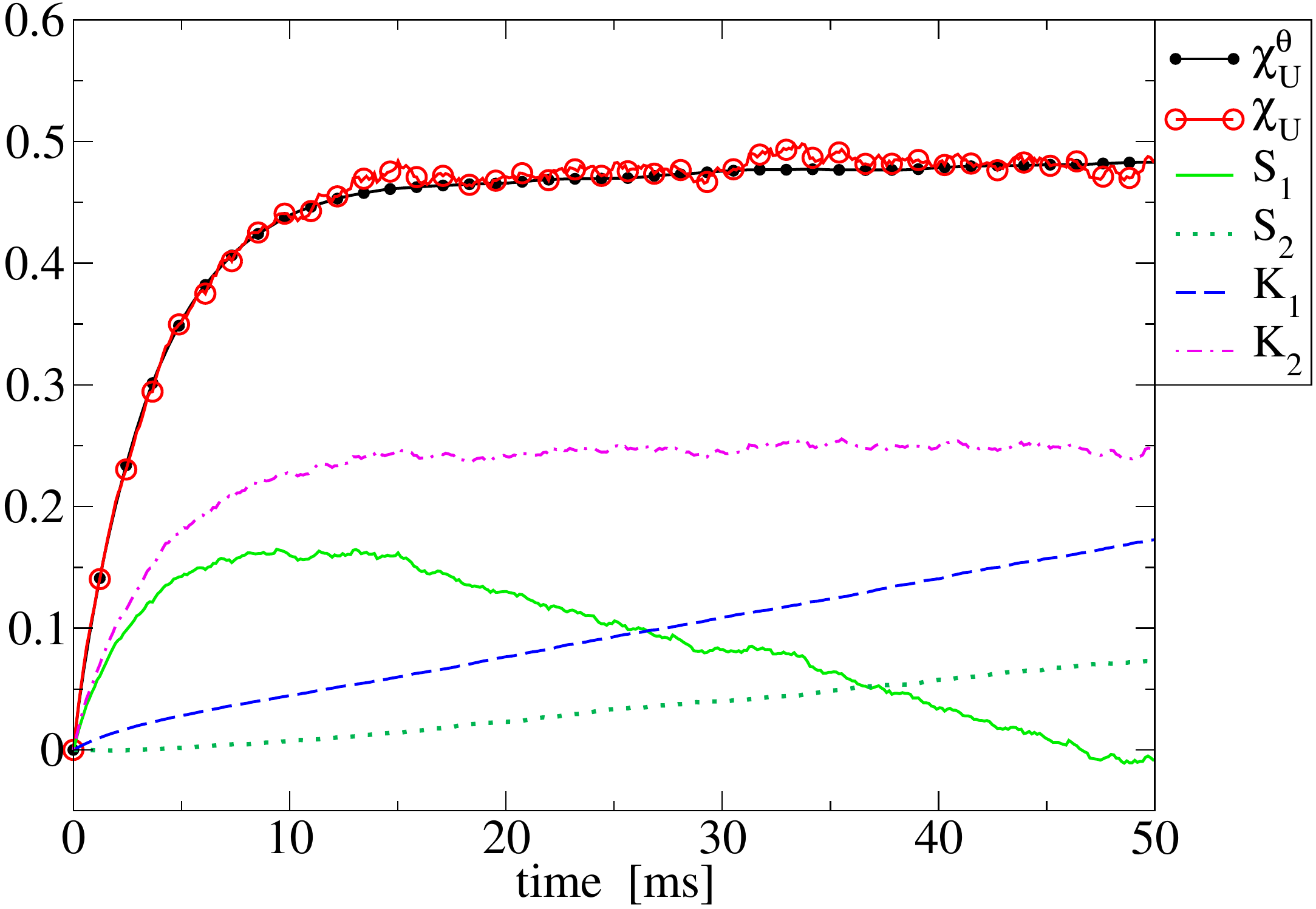}
\caption{(Color online) Response of the total energy $U$ to a change of $T_1$, for equilibrium ($T_1=296$ K $=T_2$) 
and for nonequilibrium ($T_1=140$ K and $T_1=88$ K). The susceptibility $\chi_U$ computed with
the fluctuations-response relation \eqref{chi} and its terms $S_1$, $S_2$, $K_1$, $K_2$ are shown.
The susceptibility $\chi_U^\theta$ computed with the reweighting formula \eqref{chi_rewe1} agrees with $\chi_U$. 
}
\label{fig:U}
\end{figure}

The susceptibility $\chi_U$ of the internal energy to a change of $T_1$ is
shown in Fig.~\ref{fig:U} as a function of time for the three values
of $T_1$.  It correctly converges to a constant value for large times,
though its single terms may be extensive in time in nonequilibrium conditions.
We have also an analytical argument predicting that such constant value
should be $1/2$. It is based on recently proposed 
mesoscopic virial equations \cite{falvir}.
For each degree of freedom $i$ in an
overdamped system subject to multiple reservoirs we have
\begin{align}
 - \mean{ x_i F_i (\vec x)} = T_i \,. \label{eq:virial}
\end{align}
In our system with quadratic potential energy
this implies $\mean{U(\vec x)} = T_1/2 + T_2/2$ in a
steady state. Therefore, it is expected that the susceptibility
$\chi_U (t) = \del\! \mean{U}/\del T_1 \to 1/2$ as $t\to \infty$. This is
indeed observed in the top panel of Fig.~\ref{fig:U}, where the steady
state is an equilibrium state, with $T_1=T_2$. The asymptotic value of
$1/2$ for the susceptibility is also fairly well reached by the data
in the lower panels of Fig.~\ref{fig:U}; a possible explanation of the
slight disagreement is given in the next paragraph.  In equilibrium
(top panel), $K_1$ and $S_2$ vanish while $K_2$ is equal to $S_1$,
that is, the response is given by twice $S_1$. This is essentially the
Kubo formula, stating that in equilibrium the response of an
observable is given by its correlations with the entropy produced in
the environment (heat flow divided by reservoir temperature), which is
confirmed by the form \eqref{chi_S1b} of $S_1$. (The extra
  factor of $1/T_1$ in \Eqref{chi_fr_2} has to do with the units of
  susceptibility.) On the other hand, out of equilibrium the equality
between $S_1$ and $K_2$ is lost in addition to $K_1$ and $S_2$ no
longer vanishing, as demonstrated by the two bottom panels of
Fig.\ \ref{fig:U}: All the terms $S_1, S_2, K_1,$ and $K_2$ of
\Eqref{chi_fr_2} composing $\chi_\Obs$ are all relevant. The
correlation $S_1$ between the observable and the heat flow is not
sufficient anymore in nonequilibrium systems.  
The frenetic terms $K_1$, $K_2$ and
the new entropy production term $S_2$, are also relevant for
predicting the nonequilibrium response.

%%%%%%%%%%%%%%%%%%%
%%%%%%%%%%%%%%%%%%%

While the susceptibility at equilibrium ($T_1=T_2$) attains the
expected asymptotic value of $1/2$ fairly closely, the susceptibility
out of equilibrium ($T_1 \neq T_2$) seems to fall a bit short. We
argue that this has to do with the inevitable limitation on the time
resolution of the trajectory measurements, since numerical simulations
of an equivalent system also exhibit the same feature when the time
discretization becomes coarse. Indeed, the sampling interval in the
experiments ($\approx 0.1$ ms) is not \emph{much} smaller than the dynamical
time scale $Y = 13$ms in the circuit, which one can
confirm visually from the plots in Fig.\ \ref{fig:U}. The reason why
it is the nonequilibrium susceptibilities which suffered more from
this quantization error is likely as follows: Out of equilibrium,
trajectory functionals like entropy production are numerically larger
than in equilibrium, amplifying any error in the
trajectory.

In all examples we also plot the susceptibility $ \chi^\theta_U$
computed with \eqref{chi_rewe1}. Clearly there is a very good
agreement between this estimate and  $\chi_U$ for all times, including the
deviation from the asymptotic value $1/2$ for large times.
This suggests that both approaches work well and corroborates
our explanation of the slight offset in the asymptotic value,
as also  $ \chi^\theta_U$ should be affected by the time-step discretization.

\section*{Conclusion}

We have shown that experimental steady state data can be used
to predict the thermal linear response of an electric circuit,
even if it works in a thermally unbalanced nonequilibrium regime
due to a cryogenic bath applied to one of the two resistors.
We have used a recent nonequilibrium response relation
for our analysis. This approach requires the knowledge of the
forces acting on each degree of freedom, an information easily
available in our case. The nonequilibrium version of the
heat capacity provides a simple demonstration of the fact that
in general one cannot expect to predict the response of the energy
to thermal variations just from the unperturbed correlations between
energy and fluctuating heat flows, as one would do by using
the standard fluctuation-dissipation theorem for equilibrium systems.
Also non-dissipative aspects play a crucial role:
The response includes correlations between the observable and the 
so-called frenesy of the system~\cite{bas15,bai13}, 
which is a measure of how frantically 
the system wanders about in phase space.
Eventually our example of generalised heat capacity should help understanding
how to construct a theory for steady state thermodynamics.

In order to have a comparison with an independent method for computing
the susceptibility, we also introduced a reweighting procedure that
has the advantage of needing no more than the same steady state data.
The second method estimates the susceptibility of the system in a more 
direct sense, namely mimicking actual finite perturbations of the system.
This procedure is simple to implement and is related to a linear response
formula also based on the knowledge of the steady state distribution.

\appendix
\section{Gaussian steady states distributions}

We review the procedure used to obtain the steady state distribution
for linear overdamped stochastic systems with additive noise.

Consider a process given by the stochastic
differential equation
\begin{align}
  \dot{\vec x} = -A \vec x + \sqrt{2 D} \vec \xi \ ,
\end{align}
with $\vec \xi$ being $N$-dimensional uncorrelated noise and $\vec x$
the $N$-dimensional state. Here, $A$ and $D$ are $N\times N$ positive-definite
constant matrices.
The ensemble current corresponding to these degrees of freedom follows as
\begin{align}
  \vec J = -\rho A \vec x - D \vec \nabla \rho \ ,
\end{align}
with the Fokker-Planck equation 
$\del_t \rho + \vec \nabla \cdot \vec J = 0$. 
Thus, stationarity implies (index notation hereafter)
\begin{align}
  0 =& -\del_i J_i \\
  =& \del_i (\rho A_{ij} x_j) + D_{ij} \del_i \del_j \rho \\
  =& \rho A_{ii} + A_{ij} x_j \del_i \rho + D_{ij} \del_i \del_j \rho \ .
\end{align}
Clearly an exponential quadratic form would satisfy this equation
and the ansatz
\begin{align}
\label{ss-distr}
  \rho (\vec x) = \sqrt{\frac {\det G}{(2 \pi)^N}}
  \ee^{-\frac 12 x_i G_{ij} x_j} \ ,
\end{align}
with $G$ positive-definite, yields
\begin{align}
  0 =& A_{ii} - A_{ij} x_j G_{ik} x_k 
  + D_{ij} (G_{ik} x_k G_{jl} x_l - G_{ij}) \\
  =& A_{ii} - D_{ij} G_{ij} - 
  x_k x_l ( A_{il} G_{ik} - G_{ik} D_{ij} G_{jl} ) \ .
\end{align}
Using the symmetry of $G_{ij}$ and matrix notation,
this can be rewritten as
\begin{align}
  \mathop{\rm Tr} (A - D G) 
  = \vec x^\dagger (G A -G D G) \vec x \ . \label{eq:cond}
\end{align}
Since this is supposed to hold for any $\vec x$, both sides must
vanish. For the right-hand side, this implies that the matrix 
$G A -G D G$ is skew-symmetric, which means that it 
has vanishing symmetric part,
\begin{align}
  G A + A^\dagger G = 2 G D G \ ,
\end{align}
or, equivalently, 
\begin{align}
  A G^{-1} + G^{-1} A^\dagger = 2 D \ . \label{eq:condi}
\end{align}
The left-hand side of \eqref{eq:cond} also vanishes, as required, when
a $G_{ij}$ satisfying \eqref{eq:condi} is found.

Being a linear equation in the
unknown entries of $G^{-1}$, one can imagine rewriting
\eqref{eq:condi} so as to treat those unknowns as a vector (likewise
the right-hand side), and afterwards inverting the matrix
equation. This is achieved by resorting to the Kronecker product,
denoted by $\otimes$, and a ``vectorization'' operation, denoted as
``vec'', which amounts to stacking the columns of a matrix into a
single column. Eq.~\eqref{eq:condi} is recast in the form
\begin{align}
  (I \otimes A + A \otimes I) \mathop{\rm vec} G^{-1}
  = 2 \mathop{\rm vec} D \ .
\end{align}
Hence we find $G^{-1}$ via
\begin{align}
  \mathop{\rm vec} G^{-1} = 
  2 (I \otimes A + A \otimes I)^{-1} \mathop{\rm vec} D \label{eq:vecGi}
\end{align}
followed by an ``un-vec'', \ie. a procedure reverting back from vectorized
matrices to actual ones.

\subsection*{Circuit experiments}

In the electric circuit experiments, the equation of motion for the
charges is of the form
\begin{align}
  \dot{\vec q} = -R^{-1} \Cp^{-1} \vec q + R^{-1} \sqrt{2 RT} \vec \xi \ ,
\end{align}
where 
\begin{align}
 \Cp & = \begin{bmatrix} C+C_1 & C \\ C & C+C_2 \end{bmatrix} \\
 \Cp^{-1} & = \frac 1 X \begin{bmatrix} C+C_2 & -C \\ -C & C+C_1 \end{bmatrix} \\
 R & = \begin{bmatrix} R_1 & 0 \\ 0 & R_2 \end{bmatrix} \\ 
 T & = \begin{bmatrix} T_1 & 0 \\ 0 & T_2 \end{bmatrix}
\end{align}
[$X = \det \Cp$ was defined in \eqref{X}].
Thus, we identify $A = R^{-1} \Cp^{-1}$ and 
$D = R^{-1} T$. Through \eqref{eq:vecGi} and inverting 
the resulting matrix $G$, we have
\begin{align}
  G = \frac {Y} {Z}
  \begin{bmatrix}
    g_{11} & g_{12} \\ 
    g_{21} & g_{22}
  \end{bmatrix}
\end{align}
with 
\begin{align}
  Y =& R_1R_2 (\det \Cp) (\mathop{\rm Tr}A)\nonumber\\
    =& (C+C_1)R_1 + (C+C_2)R_2 \\
  Z =& X [Y^2 T_1 T_2 + R_1 R_2 C^2 (T_1-T_2)^2]
\end{align}
and 
\begin{align}
  g_{11}=& T_2 Y (C+C_2) + (T_1-T_2)R_1C^2 \\
  g_{12}=& -(C+C_1)CR_1T_1 -(C+C_2)CR_2T_2 \\ 
  g_{21}=& g_{12}\\ 
  g_{22}=& T_1 Y (C+C_1) + (T_2-T_1)R_2C^2
\end{align}
We have thus all the elements for computing the steady state
distribution \eqref{ss-distr} analytically at any combination
of temperatures $T_1,T_2$.

%%%\bibliography{bib_noneq_20160606}


\end{document}